\documentclass[usenatbib,useAMS]{mn2e}
\usepackage{epsfig,psfig,graphicx,float,color}
\usepackage{subfigure}
\usepackage{mybibdefs}
\usepackage{multirow}
\usepackage{widetext}
\usepackage{url}
 \usepackage{color}

\newcommand{\solM}{\mathrm{M_{\odot}}}
\newcommand{\Mpc}{\mbox{Mpc}}

\newcommand{\km}{\mbox{km}}
\newcommand{\s}{\mbox{s}}
\newcommand{\h}{h}

\newcommand{\pel}{\mathrm{p_{el}}}
\newcommand{\psp}{\mathrm{p_{sp}}}

\usepackage{graphicx}
\usepackage{amssymb}
\usepackage{epstopdf}
\title[Elliptical galaxy fractions in  groups and clusters.]{The fraction of early-type galaxies in low redshift  groups and clusters of galaxies}  
\author[Hoyle et al.]{Ben  Hoyle$^{1}$, Karen. L. Masters$^{2,3}$, Robert C. Nichol$^{2,3}$, 
Raul Jimenez$^{4,1}$,  \newauthor  Steven P. Bamford$^5$. \\\\\\\\
$^1$Institut de Ciences del Cosmos, Universitat de Barcelona (ICCUB-IEEC), Marti i Franques 1, Barcelona 08024, Spain.\\
$^2$Institute of Cosmology \& Gravitation, University of Portsmouth, Dennis Sciama Building, Portsmouth, PO1 3FX, UK. \\
$^3$SEPnet, South East Physics Network, ({\tt www.sepnet.ac.uk}).\\
$^4$ICREA, Institucio Catalana de Recerca i Estudis Avancats (\tt{www.icrea.es}).\\
$^5$Centre for Astronomy and Particle Theory, The University of Nottingham, University Park, Nottingham, NG7 2RD, UK.
\\
{\tt E-mail: benhoyle1212@icc.ub.edu}
 }
  
\begin{document}
\date{Accepted ----. Received ----; in original form ----.}
\pagerange{\pageref{firstpage}--\pageref{lastpage}} \pubyear{2010}
\maketitle
\label{firstpage}
\begin{abstract}
We examine the fraction of early-type (and spiral) galaxies found in groups and clusters of galaxies as a function of dark matter halo mass. We use morphological classifications from the Galaxy Zoo project matched to halo masses from both the C4 cluster catalogue and the Yang et al (2007) group catalogue. We find that the fraction of early-type (or spiral) galaxies remains constant (changing by less than $10\%$) over three orders of magnitude in halo mass ($13 \le \log M_H/h^{-1}\solM\le 15.8$). This result is insensitive to our choice of halo mass measure, from velocity dispersions or summed optical luminosity. 
Furthermore, we consider the morphology-halo mass relations in bins of galaxy stellar mass $M^*$, and find that while the trend of constant fraction remains unchanged, the early-type fraction amongst the most massive galaxies (11$\le \log M^*/h^{-1}\solM\le$ 12) is a factor of three greater than lower mass galaxies (10$\le \log M^*/h^{-1}\solM\le$10.7). We compare our observational results with those of simulations presented in De Lucia et al (2011), as well as previous observational analyses, and semi-analytic bulge (or disc) dominated galaxies from the Millennium Simulation. We find the simulations recover similar trends as observed, but may over-predict the abundances of the most massive bulge dominated (early-type) galaxies. Our results suggest that most morphological transformation is happening on the group scale before groups merge into massive clusters. However, we show that within each halo a morphology-density relation remains: it is summing the total fraction to a self-similar scaled radius which results in a flat morphology-halo mass relationship. 
\end{abstract}
\begin{keywords}
galaxies: abundances,galaxies: clusters: general,galaxies: groups: general
\end{keywords}

\section{introduction}

 One of the central observations that must be explained by any model of
galaxy evolution is the morphology-density relation \citep{1980ApJ...236..351D}: that elliptical (or early-type,
bulge-dominated) galaxies are more commonly found in the densest
regions of the Universe, whilst spiral (or late-type, disc-dominated)
galaxies inhabit less dense regions. In hierarchical models of galaxy
formation, in an expanding universe, this observation has been
explained by the transformation of disc-dominated galaxies into bulge-dominated galaxies via {\it heirarchical merging}.  This basic picture is supported by an impressive array of 
observational evidence, matched to semi-analytic galaxy formation
models \citep[for example][]{1993MNRAS.264..201K,1996MNRAS.283.1361B,2000MNRAS.319..168C,2003ApJ...584..210G,Baldry:2003kj,2011ApJ...736...59Z}, although many questions remain over the details \citep[e.g.,][]{2010MNRAS.407.2017B,2011arXiv1109.3457H,2011MNRAS.415..811T}, and recently there has also been a growth of interest in the role internal and/or secular evolution can have on the morphological evolution, and the growth of spheroidal components in galaxies \citep[e.g.,][]{2010ApJ...714L..47O,2011ApJ...726...57C}.



The standard theoretical picture of structure formation (gravitational instability of a Gaussian initial density field) predicts that the most massive halos form at the biggest peaks of the primordial density field \citep[e.g.,][]{1984ApJ...284L...9K}, which in turn are more often found in the densest parts of the universe. This naively suggests that a morphology-density relation will results in a morphology-halo mass correlation.  In addition, recent observational evidence \citep[e.g.,][]{2011arXiv1103.0547H} suggests that the most common environmental
measures used in the literature correlate extremely well with the host
halo mass \citep[but see][who argue that nearest neighbour measures
are largely independent of halo mass]{2011arXiv1109.6328M}, so that 
the observed ``morphology-density'' relation may in fact
be a ``morphology-halo mass'' relation. Both of these arguments seem to suggest that the fraction of 
bulge-dominated galaxies in a given halo should increase with halo mass. 

However, \cite{2011arXiv1109.2599D} recently used a combination of the
Millennium Simulation  \citep[][]{2005Natur.435..629S,BoylanKolchin:2009nc} with two different semi--analytic models of galaxy formation, namely \textsc{DLB07}
\citep{2007MNRAS.375....2D} and \textsc{MORGANA}
\citep{2009MNRAS.399..827L}, and found that the fraction of
bulge-dominated galaxies in both models remains constant, as a
function of halo mass ($M_H$), over a narrow cluster mass range of 14$\le\log M_H/h^{-1}\solM\le$14.8.

 Observational evidence on the fraction of ellipticals (or early-types)
in groups and clusters of galaxies as a function of halo mass has been  limited to relatively small samples due to the difficulty in
obtaining both reliable halo masses and galaxy morphological information.
The most recent studies suggest contradictory results, probably due to the varying halo and galaxy mass scales being probed.
\cite{2009ApJ...697L.137P} agree with the De Lucia et al. (2011)
prediction, finding that the fraction of S0+elliptical (early-type) and spiral 
galaxies remains constant with halo mass (where velocity dispersion
is used as a proxy for halo mass, limited to the range
 $500<\sigma_v<1100\, \km\,\s^{-1}$). This study is based
on 72 low redshift ($0.04<z <0.07$) clusters of galaxies in the  WINGS
survey \citep{2006A&A...445..805F} that possess morphological classifications of galaxies using the photometric package \textsc{MORPHOT}, which have been trained on visually classified galaxies \citep[see][for
details]{2006A&A...445..805F}.  


    \cite{2011arXiv1110.0802C} recently studied the
morphological make-up of galaxy structures (clusters, groups and
single galaxy halos) as a function of galaxy stellar mass in different
halo mass regimes. Their sample contains 176 galaxy groups at
$0.04<z<0.1$ as well as isolated, binary and general field galaxies
over the same redshift range \citep[all from the Padova-Millenium Galaxy
Catalogue;][]{2011MNRAS.416..727C}. In addition, they add galaxies in 21
clusters from the WINGS survey \citep{2006A&A...445..805F}. This
sample spans a halo mass range of $12\le \log M_H/h^{-1}\solM\le$15, and is limited in
galaxy stellar mass ($M^*$) of $M^*>10^{10.25} h^{-1}\solM$. As in
\cite{2009ApJ...697L.137P}, the morphologies come from the MORPHOT
package. They find a {\it smooth increase/decline in the fractions of
Es-S0s/late type galaxies going from single galaxies, to binaries, to
groups} (i.e. as halo mass increases), which does not contradict the predictions
of De Lucia et al. (2011), since the latter only examine the group/cluster mass scale.

We extend the above analyses by using an order of magnitude more groups and clusters of galaxies, and by probing a range of (group and cluster) halo
masses covering $13\le\log M_H/h^{-1}\solM\le$15.8. This paper performs a single, consistent study, over a range of halo masses including both groups and clusters, and additionaly considers different stellar mass ranges, allowing us to address concerns about comparing studies with different halo and galaxy mass selections.  

To calculate distances, we assume a flat $\Lambda$CDM cosmology with
$\Omega_m, \Omega_{\Lambda},H_0=(0.3,\,0.7,\,70 \,\km\,
\s^{-1}\,\Mpc^{-1})$ and $h=H_0/100$.

Throughout this paper, we loosely refer to both groups and clusters of galaxies as just 
``clusters''. We also use the terms ``spiral'' and ``early-type'' galaxy to refer
to the Galaxy Zoo (GZ) ``clean" classifications (as discussed below), while the terms 
``bulge-dominated'' or ``disc-dominated'' refer to galaxies in the simulations divided by their bulge-disc ratio. We will associate the observed early-type class with bulge-dominated galaxies in the simulations, and likewise the GZ spiral classifications with disc-dominated galaxies in the simulations. We remind the reader that such an association may have problems as these correlations are not exact, but should be sufficient to understand the broad relationship between observations and theory.

\section{Observational and Simulated Data}
\label{data}

All of the galaxies used in this study were drawn from Sloan Digital Sky Survey \citep[see][and references therein]{York:2000gk,Gunn:2006tw,Smith:2002pca} Data Release 7 \citep[][hereafter SDSS DR7]{SDSSDR7}. We made a volume limited sample from this catalogue using cuts on the k-corrected absolute $r$ band magnitudes $M_r$, taken from the \textsc{PhotoZ} table in \textsc{CasJobs}\footnote{\url{http://casjobs.sdss.org}}\citep[][]{2008ApJ...674..768O} with $-30<M_r<-20$ and redshift of $z<0.08$.

We use Galaxy Zoo (GZ) classification probabilities to identify the galaxies as either early-type or spiral (or unclassified). The GZ project \citep{Lintott:2008ne,Lintott:2010bx}\footnote{\url{http://www.galaxyzoo.org}} used an Internet tool to enable citizen scientists to visually classify galaxies. The original version of GZ presented users with galaxies from the SDSS Main Galaxy Sample (using all galaxies which made the spectroscopic targeting criteria for the DR7 release) and asked them to be classified as either spiral or early-type (smooth). Each galaxy in this sample was classified by a median of 39 citizen scientists (with a minimum of 20). The raw results were de-biased (e.g., for the effect of higher redshift galaxies appearing smoother as  the morphological structure becomes blurred), and compared to a subset of expert classifiers in \cite{Bamford:2008ra}. The data are now publicly available \citep{Lintott:2010bx}\footnote{\url{http://data.galaxyzoo.org}}.

When selecting the early-type or spiral galaxy samples, we use the de-biased morphological results from \cite{Bamford:2008ra} who used the citizen scientist classifications to assign each galaxy with a probability of being an early-type galaxy (elliptical + S0) $\pel$, or a spiral galaxy, $\psp$. They showed that few S0's (predominantly edge-on galaxies) were classified as spiral galaxies (i.e., with high $\psp$); most S0s identified by the experts have high $\pel$ within GZ. We follow many of the GZ studies \citep[e.g.,][]{Bamford:2008ra,Schawinski:2010,Masters:2010hm,2010MNRAS.405..783M,Masters:2010rw} in using the GZ ``clean'' sample, which uses a cut of  $\psp,\pel>0.8$ to select GZ galaxies and define the early-type/spiral separation used herein. It is important to remember in interpreting our result that this selection results in large numbers of ``unclassified" galaxies (with both $\psp,\pel<0.8$) which will include the irregular galaxies which are included in blue/late-type samples of other studies. However, we note that our results are largely insensitive to the probability threshold (and therefore the number of unclassified galaxies), which we demonstrate by using the full probabilities in a weighted count (see Section 3.4). 

To enable a direct comparison with the \cite{2011arXiv1109.2599D}  simulations, we additionally impose a stellar mass $M^{*}$, cut of $M^{*}>10^{10}h^{-1}\solM$. The stellar masses were estimated from the shape of the optical (SDSS fibre) spectrum using the publicly available \textsc{VESPA} code \citep[][]{2007MNRAS.381.1252T,2009ApJS..185....1T}. At the redshifts of the GZ galaxies the SDSS fibre captures only a small fraction of the light of the galaxy. The VESPA code applies aperture corrections which should result in unbiased total stellar masses on average \citep[since as shown by][]{2003ApJ...587...55G} for a large sample of SDSS galaxies the average fibre colours are very close to the average total colours), although the error on an individual galaxy stellar mass may be quite large, and this may be a particular issue for face-on spiral galaxies which have significant colour gradients.  To examine this effect we additionally used the stellar mass estimates of \cite{Baldry:2003kj}, based on integrated SDSS magnitudes and colours. We find that our results (in Section 3.2) are insensitive to the choice of stellar mass estimate, so conclude that aperture bias errors do not critically affect our results.  

Our final galaxy sample after all these cuts consists of $N=85364$ galaxies, of which $N=28602$ (33.5\%) have $\psp>0.8$ and $N=11571$ (13.5\%) have $\pel>0.8$. 


To obtain halo masses, we used the publicly available C4 cluster catalogue \citep[][]{2005AJ....130..968M} which was applied to the SDSS DR5 spectroscopic galaxy sample \citep[][]{2007ApJS..172..634A}. The C4 algorithm identified galaxy clusters as  over-densities in seven dimensional position (3-D) and colour (4-D) space. Unlike many other cluster finding algorithms, the C4 cluster catalogue is not biased against finding clusters or groups of blue galaxies, and is over $96\%$ pure for clusters with an optical ($r$-band) luminosity of $L_r>2\times10^{11}$L$_{\odot}$. However, the sample is only 100\% complete above halo masses of $M_H=5\times10^{14}\h^{-1}\solM$, and drops dramatically to $\lesssim 30\%$ complete below $M_H=5\times10^{13}h^{-1}\solM$.


We use only those C4 clusters with a measured virial radius, and a halo mass of $M_H>10^{13}h^{-1}\solM$. We converted the virial radius and redshift to an angle subtended by the cluster on the sky $\theta_{\rm VIR}$, and the velocity dispersion to a redshift error $\Delta(z)$. We allocated cluster membership to those galaxies which were within $\pm2\Delta(z)$, and separated by an angle less than $\theta_{\rm VIR}$ from the cluster centre. In total $N=10101$ galaxies reside within $N=505$ C4 cluster halos, of which $1848$ (18\%) are early-type galaxies ($\psp>0.8$) and 2161 (21\%) are spiral galaxies ($\pel>0.8$).       
                  
We also identify galaxies in halos with masses estimated by Y07. These authors estimate the host halo masses of the SDSS DR4 galaxy sample \citep[][]{2006ApJS..162...38A} by iteratively determining the group membership based on a luminosity-scaled radius, and then estimate the halo mass from a characteristic luminosity, or a characteristic stellar mass, of the group or cluster. The mass-to-luminosity relation was calibrated from simulations \citep{2004MNRAS.350.1153Y}.  In total, $N=16567$ galaxies in our GZ sample were found to reside in $N=2140$ Y07 halos (with a mass of $M_H>10^{13}\h^{-1}\solM$), of which  3355 (20\%) are early-type galaxies ($\pel>0.8$) and 3908 (24\%) are spiral galaxies ($\psp>0.8$). 

In this study, we perform two separate analyses, first looking at the early-type (spiral) fraction among all galaxies in our GZ sample associated with a C4 cluster, and separately then performing the same analysis using all galaxies in the sample having a host halo mass estimated by Y07. 

We comment here on the possible differences between the C4 and Y07 halo catalogues. For those galaxies with both a measured C4 cluster mass, and a Y07 group mass, we find a systematic offset of $\log M_H/h^{-1}\solM=0.5\pm1.0$. This is due to  the differences in the C4 and Y07 cluster--finding algorithms resulting in single C4 clusters being ``shredded'' into multiple, smaller, Y07 groups. It is therefore difficult to perform a meaningful one--to--one comparison between C4 and Y07 halo masses as C4 halos can be associated with up to eleven Y07 groups, and on average, there are $\langle N \rangle = 2.9\pm3.5$, Y07 groups (within the C4 virial radius). In summary, it appears C4  characterises the largest halo masses, while Y07 is optimal for characterising groups of galaxies and substructures within larger (C4) systems. A detailed comparison of the two halo measurements is beyond the scope of this paper, but it is instructive to ensure our results are not dependent on the halo measurements we use.

To compare our data with models of galaxy and structure formation, we extracted $960,000$ model galaxies from the Millennium Simulation \citep{2005Natur.435..629S,2009MNRAS.398.1150B}, specifically selecting galaxies from the \textsc{DeLucia2006a\_SDSS2MASS} and \textsc{DeLucia2006a}  \citep{2007MNRAS.375....2D} tables. We chose simulated galaxies to most resemble those of the GZ galaxies by placing the following constraints on absolute r-band magnitude of $M_r\leq-20$, redshift of $z\leq0.08$,  host halo mass of $M_H\geq 10^{13}h^{-1}\solM$, and on the galaxy stellar mass of $M_\star\geq 10^{10}\h^{-1}\solM$. We identify likely early-type galaxies as those with a bulge stellar mass to total stellar mass ratio of $B/T\geq0.9$ resulting in $270,000$ galaxies (hereafter ``bulge-dominated"). Likewise, likely  disc-dominated (late-type or spiral) galaxies are those galaxies with a bulge stellar mass to total stellar mass ratio of $B/T\leq0.4$ ($250,000$). Our sample contains 44,000 clusters of mass greater than $M_H\geq 10^{13}h^{-1}\solM$, and 8681 clusters within the mass range probed by De Lucia et al. 2011. Our sample is much larger because those authors restricted their sample to the same number of clusters (100) as available from the MORGANA models, which used a smaller simulation than the Millenium Simulation \citep[see][for details]{2009MNRAS.399..827L}.  

\section{Results}
\label{results}
The fraction of elliptical (or more precisely bulge-dominated) galaxies in simulations has recently been shown to be constant over the range of cluster halo masses of 14$\le \log M_H/h^{-1}\solM\le$14.8  \citep{2011arXiv1109.2599D}.  This result was obtained by identifying bulge-dominated model galaxies from an N-body simulation. De Lucia et al. (2011) selected $100$ dark matter cluster halos and followed the merger histories of their component galactic size dark matter sub-halos. They used sophisticated semi-analytic modelling of the bulge and disc, and gas cooling flows, to determine the present day ($z=0$) bulge to disc ratio of the galaxies, and identified a galaxy as an early-type if its bulge mass/total mass $B/T>0.9$ (also our selection above). These results reproduced those of previous observations by \cite{2009ApJ...697L.137P}, which were based on the automated morphological classifications of a few hundred galaxies in $77$ galaxy groups and clusters. 

\subsection{The dependence of morphology on halo mass}
\begin{figure}
   \centering
  \includegraphics[scale=0.4, clip=true, trim=35 15 15 35]{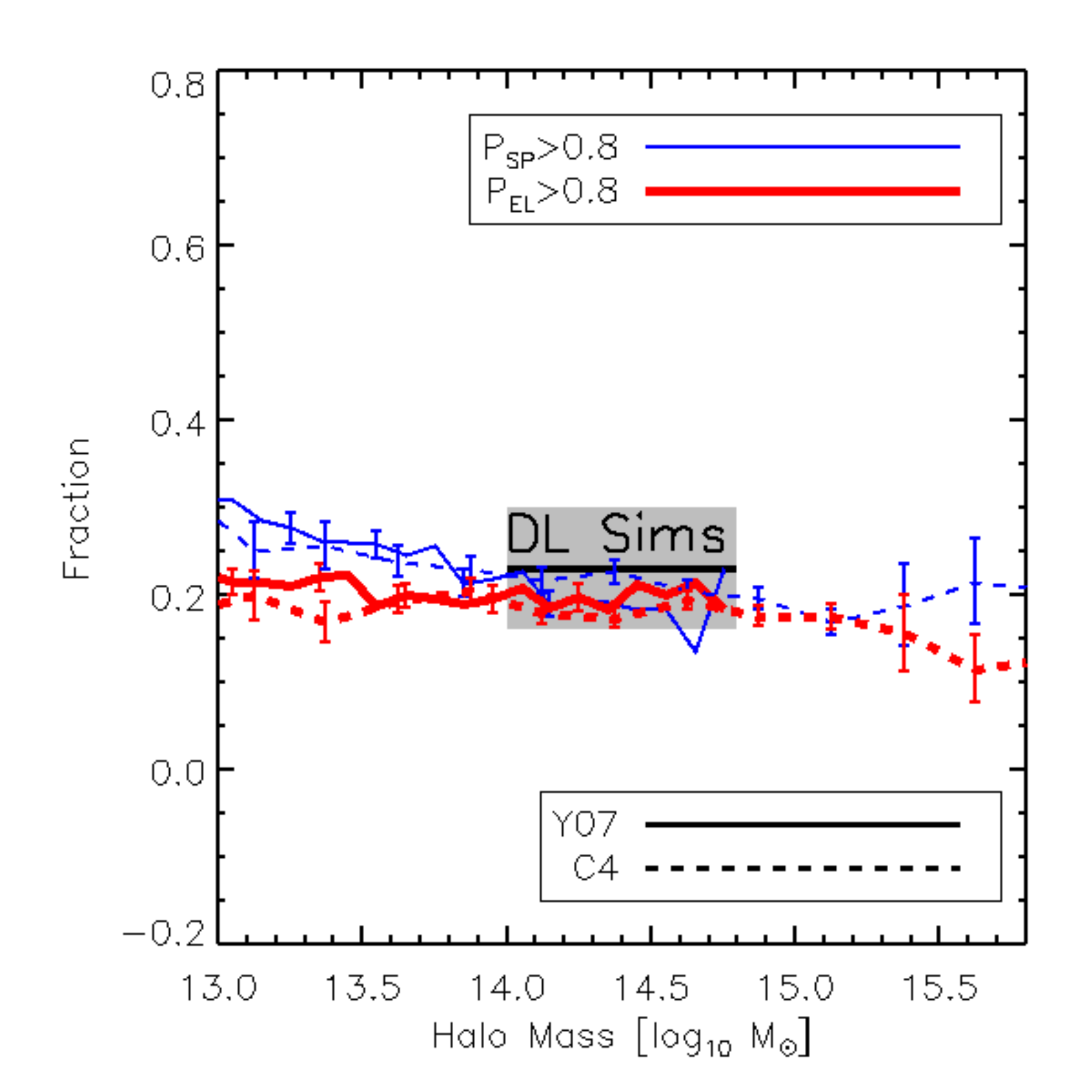} 
   \caption{ \label{f3a} The fractions of early-type (thick red lines) and spiral galaxies (thin blue lines) per halo with respect to the total number of galaxies per halo, as a function of halo mass. We show the Y07 halo mass estimates as the solid lines, and the C4 cluster halo estimates using the dashed lines. The error bars are Poisson errors. The grey shaded region (labelled ``DL Sims'') indicates a prediction for the  early-type galaxy fraction obtained by simulations (De Lucia et al. 2011) which span the mass range 14$\le \log M_H/h^{-1}\solM\le$14.8.}
\end{figure}

Our main result is that we agree that the fraction of early-type galaxies in a halo is insensitive to halo mass. We show this is in Fig. \ref{f3a} where we plot the fraction of early-type (and spiral) galaxies in a halo (as a function of halo mass for both the C4 and Y07 halo catalogues). The error bars are Poisson counting errors \citep[a good approximation of the errors on fractional quantities in large samples which are not close to either zero or 100\%,][]{Cameron:2010he}. We also plot the result of the simulations by \cite{2011arXiv1109.2599D} using a constant value for the elliptical fraction, $f_E =0.24\pm 0.07$, which is the average result of the two models presented by them, and our estimate of the $1\sigma$ dispersion in their models (from their Fig. 1). The shaded region indicates both the range of halo mass in their model and our estimate of the dispersion. 

We remind the reader that the fraction of early-type and spiral galaxies from GZ, add up to less than one; some galaxies have neither $\pel>0.8$ or $\psp>0.8$. We find that the fraction of early-type galaxies is independent of halo mass for the C4 catalogue halo mass estimates. This result has been seen before in Fig. 13 of \cite{Bamford:2008ra} using the same GZ classifications and the C4 cluster catalogue. We extend this previous analysis by also showing the independence on halo mass also holds using the Y07 halo masses.  In addition, the fraction of spiral galaxies is shown to decline smoothly from 0.3 to 0.2  over the same mass range (again using either Y07 or C4 clusters catalogues), and that this process begins from halos with masses as small as 13$\log M_H/h^{-1}\solM$.


\subsection{The dependence of morphology on stellar mass}
\begin{figure}
   \centering
   \includegraphics[scale=0.4, clip=true, trim=35 15 15 35]{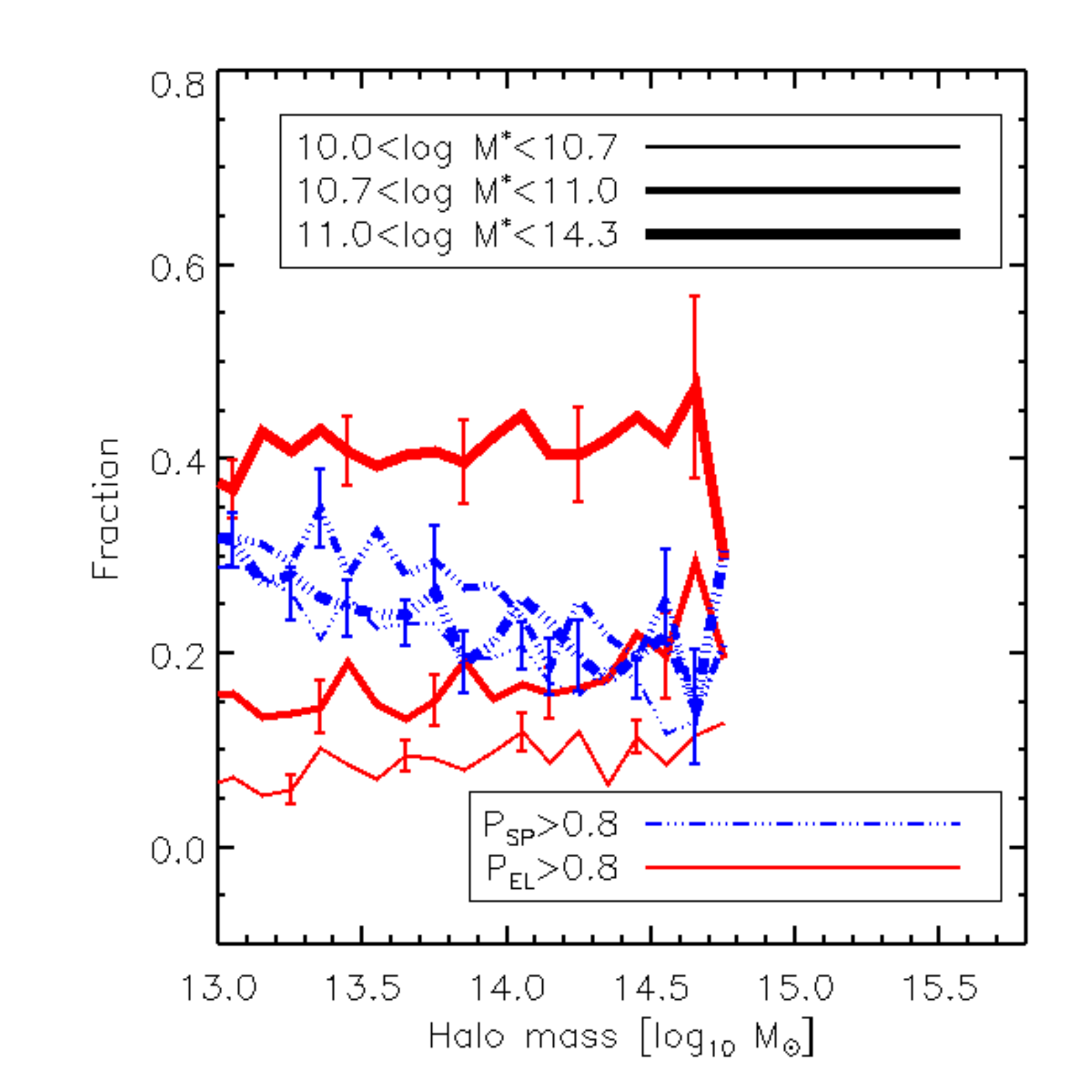} 
      \includegraphics[scale=0.4, clip=true, trim=35 15 15 35]{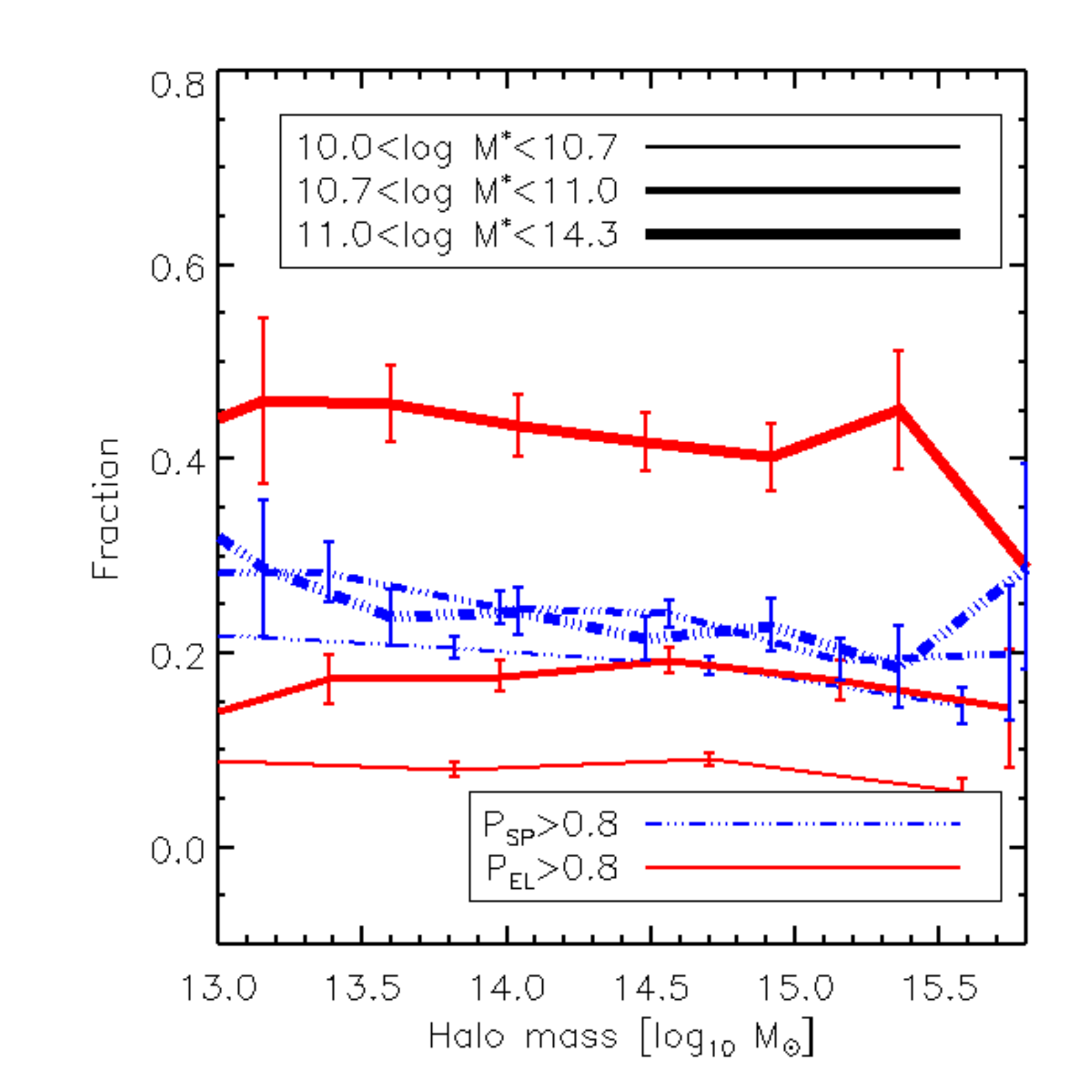} 
   \caption{ \label{f5} The morphological fractions of galaxy as a function of  halo mass using the Y07 (C4) catalogue, separated into sub-samples of stellar mass in the upper (lower) panel. Here the early-type galaxies are shown by the red solid lines, and the spiral galaxies by the blue dot-dashed lines. The increasing  line thickness corresponds to increasing bins of stellar mass, and errors are Poisson, of which we show every third (all) for clarity. }
\end{figure}

We next examine the fraction of early-type and spiral galaxies as a function of both halo mass and galaxy stellar mass. We divide the galaxy sample into bins of total stellar mass M$^*$ and we compute the fraction of early-type, and spiral, galaxies in each bin with respect to the total number of galaxies, as a function of halo mass ($M_H$). In the top panel of Fig. \ref{f5}, we show these results using the Y07 halo catalogue, and in the bottom panel we show the results using the  C4 cluster catalogue. The width of stellar mass bins were chosen to contain approximately equal numbers of galaxies.

In Fig. \ref{f5}, we see that the fraction of early-type galaxies in a halo remains constant as a function of halo mass (within the errors) within each stellar mass bin. However, we see that the fraction of early-type galaxies per halo increasing by a factor of $2$ to $3$ as we increase the galaxy stellar mass over the range 10.5$\lesssim \log M^*/h^{-1}\solM\lesssim$11.5. It is however unclear how much this affect is due to more massive, and therefore luminous, galaxies being easier to classify, so the fraction of unclassified GZ galaxies which should be classified as early-type may drop in the high stellar mass bins. We note that the fraction of unclassified galaxies drops as a function of increasing stellar mass.

 In contrast, we see that the fraction of spiral galaxies in a halo is largely unaffected by either stellar mass, or halo mass. We note that we have also examined the effect of splitting the galaxy sample into bins of total galaxy metallicity (as derived by \textsc{VESPA}), but find no correlation between metallicity and changes to the fraction of either early-type, or spiral galaxies. 

Recently \cite{2011arXiv1110.0802C} also found that the fraction of elliptical (and the combination elliptical+S0) galaxies increases as a function of increasing stellar mass, but they find that the fraction of late-type galaxies decreases as a function of stellar mass. We note that the late-type galaxies of  \cite{2011arXiv1110.0802C}  are defined as containing both spiral and irregular morphological types, whereas our work has concentrated on specifically spiral galaxies from the GZ classifications. Particularly at low stellar masses, most late-types may be irregulars \citep[e.g.,][]{1988ARA&A..26..509B} which would likely be ``unclassified'' in GZ.
 
\subsection{Comparisons with simulations}
\begin{figure}
   \centering
   \includegraphics[scale=0.4, clip=true, trim=35 15 15 35]{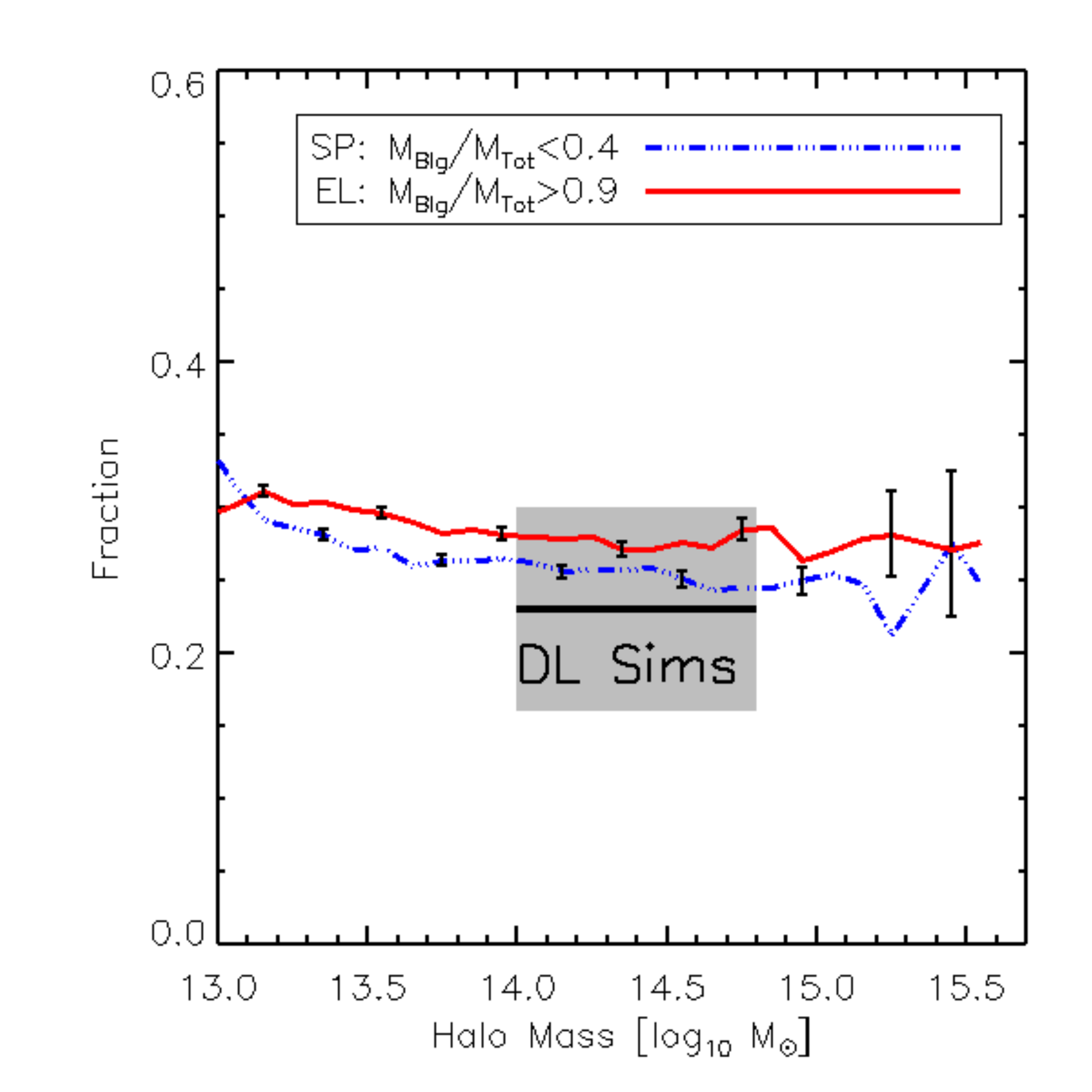} 
   \includegraphics[scale=0.4, clip=true, trim=35 15 15 35]{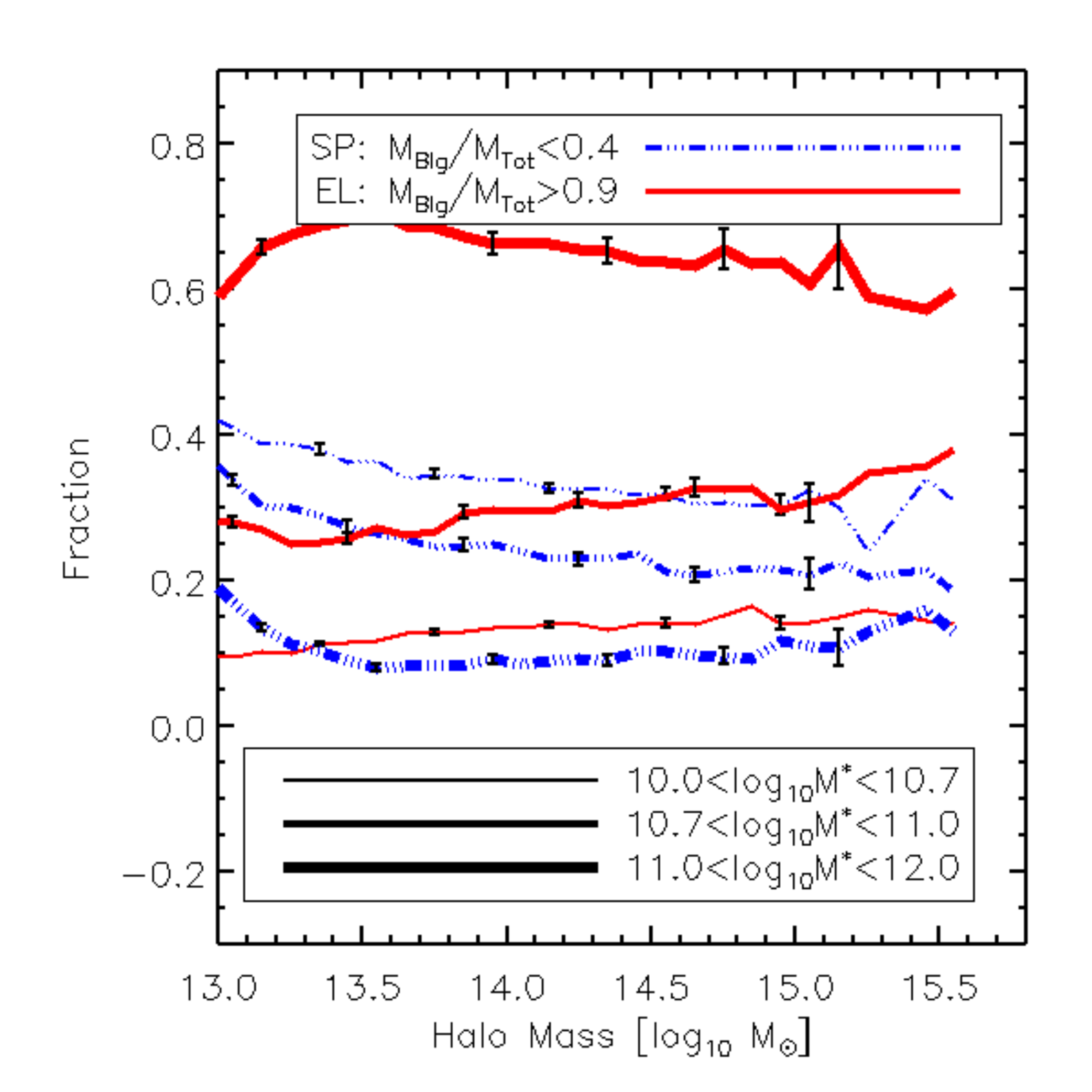} 
   \caption{ \label{fss} The fraction of bulge-dominated (early-type) galaxies (red solid lines),  and disc-dominated galaxies (blue dot-dashed lines) as a function of halo mass, drawn from the Millennium simulations. The upper panel shows the ratio of the number of bulge (disc) dominated galaxies at each halo mass, to the total number of galaxies at each halo mass.  We plot the results of the more detailed semi-analytic simulations from De Lucia et al. (2011) as a shaded region. In the lower panel, we divide the galaxy samples into bins of total stellar mass, the thicker lines correspond to bins of greater stellar mass as indicated in the key. All error bars are twice Poisson to help visualise their size. }
\end{figure}

We next present the fraction of bulge-dominated (or early-type) galaxies, and disc-dominated (or spiral) galaxies, drawn from the Millennium Simulations, as a function of halo masses. In the upper panel of Fig. \ref{fss}, we show the ratio of the number of bulge (disc) dominated galaxies at each halo mass, to the total number of galaxies at each halo mass.  We again plot the results of the more detailed semi-analytic simulations from \cite{2011arXiv1109.2599D} as the shaded region. In the lower panel, we split the galaxy samples into bins of total stellar mass, with the thicker lines depicting bins of increasing stellar mass. We show the error bars as twice the Poisson error so they are visible.

Our results show that the fraction of bulge-dominated (or early-type)  and disc-dominated (or spiral) galaxies in the Millennium Simulation is relatively constant as a function of halo mass above  $\log M_H / h^{-1}\solM = 13.2 $, changing by at most $10\%$ over two orders of magnitude in $M_H$ (Fig. 3).  We find that the bulge-dominated galaxy fraction is in good agreement with the results  presented in \cite{2011arXiv1109.2599D}, who used a smaller sample size. With these results however, we can extend the De Lucia et al. (2011) findings over a wider range of halo masses now showing that the fraction of elliptical (or more correctly, bulge-dominated) simulated galaxies are independent of host halo mass over three orders of magnitude. 
 
The lower panel of  Fig. \ref{fss} shows how the fractions of bulge-dominated (early-type) model galaxies change as a function of halo mass for galaxies in different stellar mass bins. We find that the fraction of the most massive bulge-dominated galaxies increases by a factor of $3$--$6$ compared with the lower mass  bulge-dominated galaxies. These trends are in qualitative agreement with the data (as seen in both panels of Fig. \ref{f5}). We do however find that the fraction of the most massive bulge-dominated (early-type) galaxies is 50\% higher than found by observations. The simulation seem to over produce the number of massive  bulge-dominated (early-type) galaxies.  However, as noted above, the large fraction of unclassified galaxies in GZ (i.e., neither early-type or spiral) may introduce an observational  bias, because  more massive galaxies may be easier to classify. We note here that the total fraction of bulge-dominated simulated galaxies (28\%) in massive halos is also greater than the total number of GZ early-type galaxies in massive halos (18\% for C4, 20\% for Y07).

The disc-dominated model galaxies in the top panel of Fig. \ref{fss}, show a fractional change of less than $10\%$ over 13$\le \log M_H/h^{-1}\solM\le$15 in agreement with our observations. The same mild trend is still apparent if we divide the disc-dominated galaxies by stellar mass (see the lower panel of Fig. \ref{fss}). However, contrary to the data, the total abundances of disc-dominated galaxies changes as the stellar mass increases. This could be real or could be caused by the loose comparison of disc-dominated simulated galaxies to observed spiral galaxies, or the relatively large errors in stellar mass estimates. 

\subsection{Using the full GZ probabilities}
\begin{figure}
   \centering
  \includegraphics[scale=0.4, clip=true, trim=35 15 15 35]{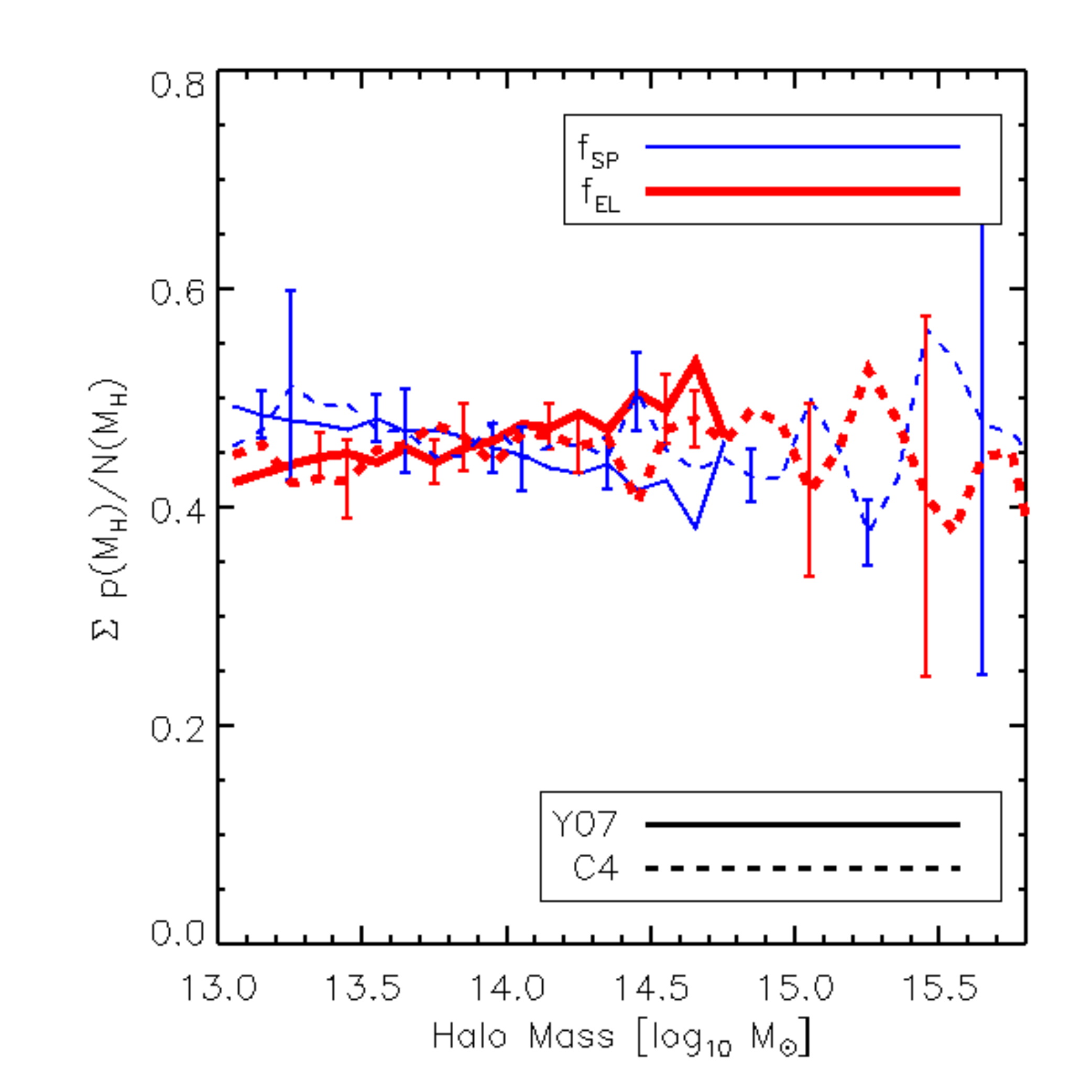}
   \caption{ \label{f33a} The probability weighted fractions (see text) of early-type (thick red  lines) and spiral galaxies (thin blue lines) with respect to the total number of galaxies per halo, as a function  of halo mass.  We show the Y07 halo mass estimates by the solid lines, and the C4 cluster halo estimates by the dashed lines. Poisson error bars are shown.}
\end{figure}

Rather than imposing a strict morphological probability cut on our galaxy sample (which results in a significant fraction of unclassified GZ galaxies), in this section we instead use the full galaxy sample, weighting each galaxy by the morphological probability assigned by GZ. Thus, we calculate the sum of the probabilities divided by the total number of galaxies per halo, as a function of halo mass, i.e., $f_{\rm el} = \sum \pel(M_H) /N(M_H)$ and  $f_{\rm sp} = \sum \psp(M_H) /N(M_H)$. In Fig. \ref{f33a}, we show the outcome of this analysis, and use the same line styles and colours as in  Fig. \ref{f3a}.  In agreement with Fig. \ref{f3a}, we again observe that the fraction of early-type (and spiral) galaxies is independent of halo mass. 

Considering in more detail the comparison between Fig. \ref{f33a} to Fig. 1 we see that the fraction of both early-type and spiral galaxies has increased: as expected since we are now including all galaxies, not only a subset with $\pel,\psp > 0.8$.  We see that the morphological fractions for both the C4 halo masses and Y07 halo masses are consistent with each other, differing by only 0.35 sigma for the early-type galaxies (and 0.3 sigma for the spiral galaxies). The fractions of early-type and spiral galaxies remain relatively flat as a function of the halo mass, when using the C4 catalogue, and change by less than  $20\%$ for the Y07 halo masses. Both early-type galaxy distributions are consistent with a constant fraction of 0.46, i.e., are independent of halo mass, with a $\chi^2$ value of 0.5 (0.6) for the Y07 (C4) catalogues.

We therefore conclude that our results are relatively insensitive to the choice of classification probability threshold used to identify early-type and spiral galaxies from GZ. 

\section{Discussion}
\label{diss}
The most intriguing result of De Lucia et al. (2011) and our above result, is that the  fraction of early-type galaxies is constant with halo mass, if we integrate the total number of galaxies in the cluster out to the virial radius.  This may appear to be in contradiction with the strong morphology-density relation (e.g., seen in \cite{1980ApJ...236..351D}, and using GZ by Bamford et al. 2009 and \cite{2009MNRAS.399..966S}), which finds that there are more early-type galaxies in denser regions of the Universe.

\begin{figure}
   \centering
   \includegraphics[scale=0.4, clip=true, trim=35 15 15 35]{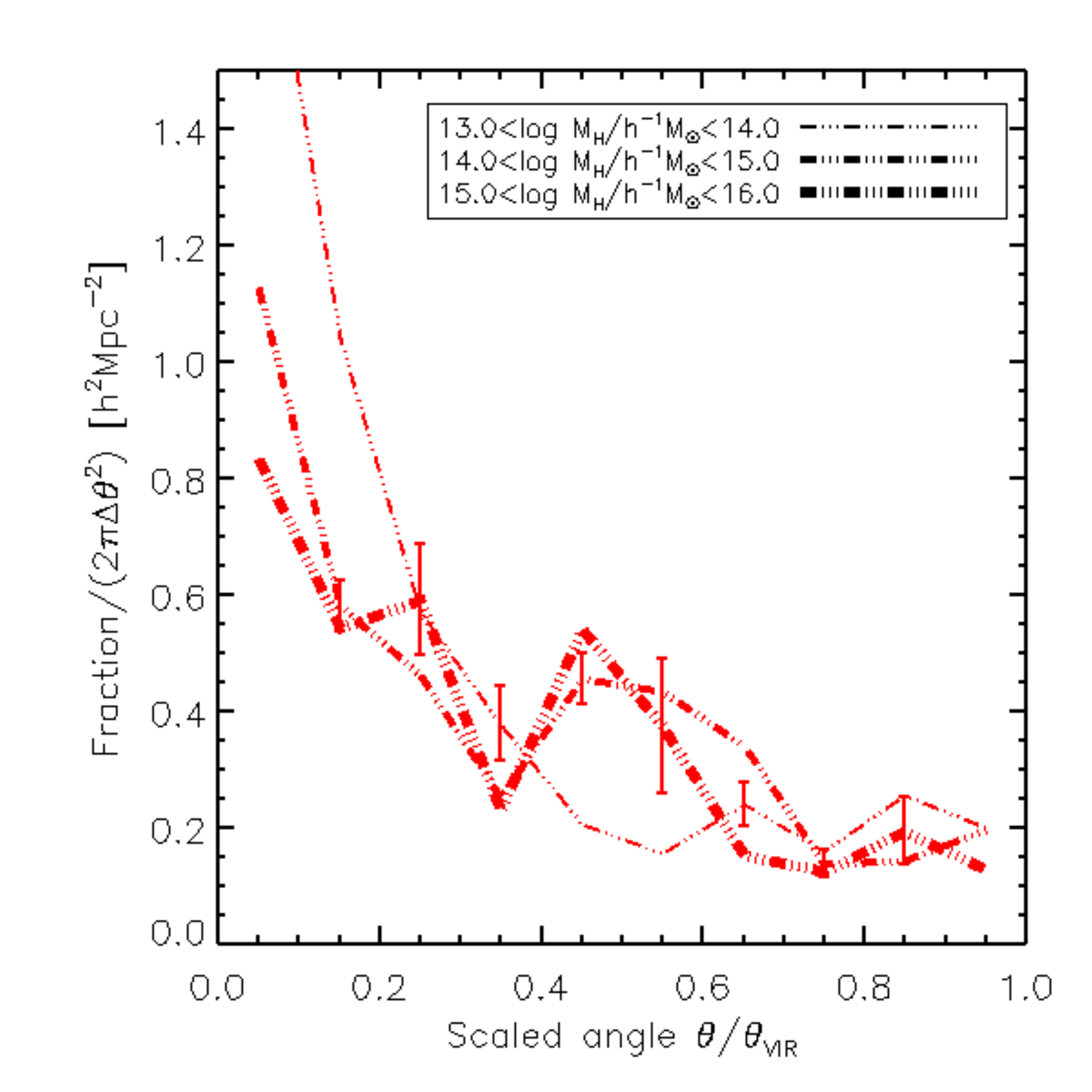} 
   \caption{ \label{cc1} The fraction of early-type galaxy as a function of scaled angular separation from the centre of C4 halos. Fractions are divided by the volume contained in the scaled annulus to allow comparison between different sized halos.  The increasing line thickness correspond to increasing bins of halo mass, and the errors (are Poisson) are shown for a representative sample of points. }
\end{figure}

We can explicitly examine the morphology-density relation within the cluster virial radius using C4 catalogue. We examine the fractions of early-type galaxies internal to a halo as a function of the radial distance from the centre of each cluster (which we scale to the virial radius).  We calculate the fraction of early-type galaxies as a function of angular separation from the cluster centre $\theta$, normalised to the angular extent of the cluster on the sky $\theta_{VIR}$. To increase signal-to-noise we stack clusters in bins of halo mass. In Fig. \ref{cc1} we show the fraction of early-type galaxies in each annulus divided by the area of the scaled annulus, as a function of scaled radius for different bins of halo mass.

We find that this scaled fraction of early-type galaxies increases towards the centre of the cluster, reproducing the strong morphology-density relation seen previously. The relationship is insensitive to the halo mass bin, due to the scaling by virial radius, effectively removing the mass dependence through the mass radius relation. We have examined the sensitivity of this result by using an alternative mass-radius relationship \citep[from][]{2007arXiv0709.1159J}, and find the observed morphology-density relationship to remain unchanged. 

So, when we consider the total fraction of early-types in a halo out to the virial radius we find no dependence on halo mass; but looking inside the halo we recover the morphology-density relation. Put together this demonstrates the self-similar nature of the number density profile of galaxies in massive halos; by summing the fraction out to the virial radius we always end up with the same total fraction of early-types, independent of the halo mass. This is not in contrast to any morphology-density relation because by definition the average density out to the virial radius of a group or cluster (or any virialised structure in the Universe at a given redshift) is the same independent of the total mass of the structure (as explained in detail in Poggianti et al. 2010). Our results support and extend the findings of \cite{2010MNRAS.405..995P}, who perform a joint analysis of simulations drawn from the Millennium simulation, and observations of 227 Abell clusters with SDSS spectra \citep[see][for details]{2006ApJ...642..188P}, and find a constant number density of star-forming galaxies as a function of halo mass.

\section{conclusions}
\label{conclus}

Using a sample of galaxies with visual morphological classifications from the Galaxy Zoo project (Lintott et al. 2008,2011), combined with halo mass estimates from the Y07 group catalogue \citep[][]{2007ApJ...671..153Y} and C4 cluster catalogue (Miller et al. 2005) we find:

\begin{itemize}
\item The fraction of early-type galaxies in a given halo (defined as having $\pel>0.8$ from GZ classifications) remains constant over a large range of halo mass (13$\le \log M_H/h^{-1}\solM\le$15.8). This observation holds using either halo masses from the Y07 group catalogue or the C4 cluster catalogue \citep[also see Figure 13 of][]{Bamford:2008ra}, and is insensitive to the exact details of the classification probability used to identify early-types in GZ. As shown in \S\ref{diss}, this is not in contradiction with the morphology-density relation.
\item The fraction of spiral galaxies in a halo (defined as $\psp>0.8$ from GZ classifications) evolves slowly ($< 10\%$ for both  the C4 and Y07 clusters) as a function of halo mass.
\end{itemize}

We next divided the galaxy sample into bins of galaxy stellar mass $M^*$, and find constant fractions of early-type galaxies (or slowly changing fractions of spirals) as a function of halo mass in each stellar mass bin.
However, we find a larger fraction of early-type galaxies (by a factor of three) amongst the most massive galaxies (11$\le \log M^*/h^{-1}\solM\le$12) in a halo than are seen amongst lower mass galaxies  (10$\le \log M^*/h^{-1}\solM\le$10.7). The spiral fraction in group and clusters size halos appears to be insensitive to the stellar mass range of the galaxies considered. 

We compare the above observational results to the \cite{2011arXiv1109.2599D} simulations as well as galaxies and  halos selected from the Millennium Simulation, using the same redshift, mass, and absolute magnitude selection criteria as the data. We identify the early-type galaxies in the MS as those galaxies which are bulge-dominated, with a bulge stellar mass to total stellar mass ratio of $B/T>0.9$. 

Our main results from this analysis are:
\begin{itemize}
\item The fraction of bulge-dominated (early-type) galaxies in the simulations is constant (to within 5\%) as a function of halo mass over the range 13$\le \log M_H/h^{-1}\solM\le$15.6.
This result is in good agreement with the observations above, and with the results of \cite{2011arXiv1109.2599D},  who used a smaller sample of clusters to select bulge-dominated (early-type galaxies) in cluster halos (over a smaller mass range of 14$\le \log M_H/h^{-1}\solM\le$14.8). We  can extend their prediction over a wider range of halo mass (13$\le \log M_H/h^{-1}\solM\le$15.6) .
\item The fraction of disc-dominated galaxies in the simulations drops from 0.3 to 0.2 as the halo mass increases from 13$\le \log M_H/h^{-1}\solM\le$15.6, and that this process begins from halos with masses as small as 13$\log M_H/h^{-1}\solM$. These trends are observed for the GZ spiral galaxies in both the C4 and Y07 halos.
\end{itemize}

We again divided the model galaxy sample into bins of galaxy stellar mass $M^*$ and we found that bulge-dominated galaxies in the simulations  are more common amongst the most massive galaxies  (similar to our observed result) but the difference between the fraction of bulge-dominated (early-type) galaxies amongst the most massive and least massive galaxies is $\sim 50\%$ greater than we observed. This result may, however, be explained by the large fraction of unclassified galaxies in the GZ sample, as more luminous, massive galaxies in our volume limited sample, are larger, and easier, to classify morphologically. The disc-dominated galaxies in the simulations are more common (by a factor of $3$) amongst the least massive galaxies in the halos than they are amongst the most massive. This is contrary to what we observed (that in group and cluster sized halos there was no change in GZ spiral fraction with stellar mass), but agrees with the observations of \cite{2011arXiv1110.0802C} who found that the the most massive late-type (disc-dominated) galaxies are less abundant, by a factor of two, in clusters, compared with groups.

The flatness of the fraction of early-type (or S0+elliptical) and spiral galaxies with halo mass has been seen before, albeit over a smaller range in mass (or velocity dispersion $500<\sigma_v<1100\, \km\,\s^{-1}$) by \cite{2009ApJ...697L.137P}, using 72 low redshift ($0.04<z<0.07$) clusters from the WINGS \citep{2006A&A...445..805F} survey. We extend this analyses by using an order of magnitude more clusters, and by probing a far greater range in velocity dispersion (equivalent to $100 \le \sigma_v \le4000\,\km\,\s^{-1}$). Our results are also in general agreement with the recent paper by \cite{2011arXiv1110.0802C}, who show that the fraction of ellipticals galaxies remains relatively flat (at $\sim0.3$) between groups and clusters.

Furthermore, after the submission of this paper, \cite{2012ApJ...746..160W} examine the morphology density relation, using 911 galaxies with morphological classifications from RC3 \citep[][]{1991trcb.book.....D} with further checks  for morphological type using with SDSS imagining, (note that 406 of these had a dubious classification after visual inspection, of which 165 were later reclassified) within 729 Y07 halos of mass 11$\le \log M_H/h^{-1}\solM\le$14.8. In that study most halos have only one galaxy, and they find a strong dependence of morphology on halo mass, in particular in the fraction of S0 galaxies, a classification we are unable to probe with GZ1. For comparison, our study is larger, and uses 2140  halos with 3355 ($\pel>0.8$) early-type and  3908 ($\psp>0.8$) spirals galaxies.

It is fascinating that while there is a strong morphology-density relation (seen above and in Dressler et al. 1980, and in GZ Bamford et al. 2009 and Skibba et al. 2009) that between group and cluster sizes halos we see no (or little) change in the early-type (or spiral) galaxy fraction per halo. The interpretation of this result must be that the transformation between spiral and early-type galaxies is predominantly happening on the galaxy group scale, and very massive galaxy clusters are consistent (at least in this observation) with having been made simply from a hierarchical aggregation of galaxy groups with no further need for morphological transformation of the galaxies inside the groups.  That processes which transform galaxy morphologies appear to happen preferentially on the galaxy group scale has been discussed before \citep[][]{1998ApJ...496...39Z,2001ApJ...562L...9K,2003MNRAS.339L..29H, 2008ApJ...688L...5K}. Our observations and extension of the mass range of the semi-analytical models of De Lucia et al. (2011) support this basic picture, and suggest that it is well incorporated into basic semi-analytic models of galaxy formation in an hierarchical universe.  


\section*{Acknowledgments} 
\label{ack}
BH  acknowledges  grant number FP7-PEOPLE- 2007- 4-3-IRG n 20218 and the ICG for hospitality, and KLM acknowledges funding from The Leverhulme Trust as a 2010 Early Career Fellow. RCN acknowledges STFC Rolling Grant ST/I001204/1 to ICG for ``Survey Cosmology and Astrophysics".  SB acknowledges receipt of an STFC Advanced Fellowship.
Funding
for the SDSS and SDSS-II has been provided by the Alfred
P. Sloan Foundation, the Participating Institutions, the
National Science Foundation, the U.S. Department of
Energy, the National Aeronautics and Space Administration,
the Japanese Monbukagakusho, the Max Planck
Society, and the Higher Education Funding Council for
England. The SDSS Web Site is http://www.sdss.org/. 
This publication has been made possible by the participation of more than 160 000 volunteers in the Galaxy Zoo project. Their contributions are individually acknowledged at http://www.galaxyzoo.org/Volunteers.aspx.

\bibliographystyle{mn2e}
\bibliography{clustellip}

\end{document}